# A Survey on Conceptual model of Enterprise ontology

**Zeinab Rajabi,** Assistant Professor, Faculty of Computer Engeenering Department, Hazrat-e Masoumeh University, Qom, Iran, z.rajabi@hmu.ac.ir(corresponding author)
**Seyed Mohsen Rahnamafard**, University of Tehran,Tehran, Iran, rahnamafard@alumni.ut.ac.ir

**Abstract**:
Enterprise ontology serves as a foundational framework for semantically comprehending the nature of organizations and the essential components that uphold their integrity. The systematic and conceptual understanding of organizations has garnered significant attention from researchers due to its pivotal role in various domains, including business modeling, enterprise architecture, business process management, context-aware systems, application development, interoperability across diverse systems and platforms, knowledge management, organizational learning and innovation, and conflict resolution within organizations. Achieving a consensus on the concepts related to the fundamental elements that constitute an organization is therefore critical.

This study aims to conduct a comprehensive analysis and comparison of existing conceptual models of enterprises as documented in scholarly articles published over the past decade. We discuss the strengths and weaknesses of each model and introduce a robust framework for their evaluation. To facilitate this evaluation, we propose several pertinent criteria derived from established methodologies for assessing ontologies. Furthermore, we identify contemporary challenges and issues that have been overlooked in prior studies, offering insights and suggestions for future research directions in enterprise modeling. This article ultimately presents a roadmap for enhancing the systematic understanding of organizations through refined enterprise ontology frameworks.

**1-Introduction**

Throughout the literature, scholars have provided numerous definitions for the term organization. Most define an organization as a group of individuals who have come together to achieve a specific goal[1] . An organization is a phenomenon whose most important constituent is human beings. According to postmodernist beliefs shared by several philosophers over the past couple of decades, enhancing interactions among humans is one of the most important areas for achieving high-performance organizations[2]. One relevant topic in this area is semiotics, which concerns interactions formed through signage and sign processes, including modeling languages[3].

On the other hand, according to systems theory, an organization is also a kind of one of the many systems around us. Most scholars whose ideas are based on Newtonianism and systematic thinking believe that an organization is a system consisting of elements between which distinct relationships exist; thus, understanding organizations depends on grasping these elements and their relationships[4]. Despite these scientific approaches, there is still a lack of a common language that would help bring about a general consensus in the research community[5].

Management scholars have developed models to better understand organizations, their organizational problems, and to help classify and interpret their data systematically; this was presented by Rock and Crawford. Some studies such as [6],[7],[8],[9],[10] and model the relationships between organizational components. For example, Leavitt[11], in his diagnostic organizational model, introduces organizational components such as structure, technology, people, and tasks. Weisbord presents a six-box organizational cognitive model[12], [9] , which includes purposes, structure, relationships, rewards, leadership, and helpful mechanisms. Janicijevic[13] examines and compares the organizational cognitive model, dividing organizational elements into two categories: static and dynamic. Static elements include organizational structure, systems, culture, informal groups, and power structure, whereas dynamic elements are business processes, group processes, leadership, conflicts, political processes, and communication. After reviewing the literature, he concludes that existing diagnostic organizational models are imperfect because they do not include dynamic formal organizational components such as business processes. According to him, a complete and comprehensive diagnostic model should encompass business processes and related parameters such as process owners and participants, organizational competence, key performance indicators, business process shortcomings and problems, key paths to change business processes, and business process priorities.

On the other hand, information technology researchers want to align information technology with the goals of the organization and also develop information systems after the organization's structured recognition. Artificial intelligence researchers also seek to understand the structure of organizations with the aim of creating a suitable environment for their own systems. Ontology is the most relevant area of science that aims to recognize these aspects of organizations. This field was first introduced by artificial intelligence experts in order to make sense of human semantic treasure for machines. The field of ontology has developed methods and tools to build different conceptual models for verbal and non-verbal concepts in various subject domains, one of which includes that of organizations. Roseing[14] reviewed business ontology research and examined how business ontology is used in organizational development. He used the potential of ontology and semantics to develop standards that describe objects, relationships, and rules for enterprise modeling, organizational engineering, and enterprise architecture.

Enterprise ontology provides a uniform representation of similar semantic content[15]. Modellers use different methods to develop models. These models are created with different languages and modeling tools. There may even be various styles and different techniques used within a single method. In addition to this, products created by different organizations and disciplines use different terms to analyze organizations, which leads to various perceptions of the organization. Therefore, a standardized format is needed to translate data among different systems of the organization and to understand different models of organizational analysis[16].

Enterprise ontology provides a data structure that facilitates the reader's understanding of data usage in an organization description document. For example, Rajabi et al. [17]

present the methodology for enterprise architecture development based on enterprise ontology. The ontology of the enterprise provides the necessary information to collect, organize, and store data in an easy way to understand[18] [19]. For example, the Dodaf Data Meta Model[20] states that the goal of a conceptual model is to support the integrity and semantic accuracy of architectural descriptions.

On the other hand, enterprise ontology helps to model more efficiently by describing enterprise building blocks and their relationships. The enterprise ontology is a proper basis for an integrated understanding of an organization's elements. The enterprise ontology actually models the building blocks of organizations with their relationships according to the perception of entities from two parties[16]. The relationships among all elements of the organization are modeled precisely, transparently**,** and are formulated in the ontology of the organization; then a common model is created that has the necessary precision for all parties within the organization and systems.

It is quite clear for researchers the advantages and successful applications of ontology in business and various applications[21]. Ontology development for organizations is the proper basis for enterprise architecture methods[17],[19],[22] automatic analysis of models at enterprise architecture, querying and inference in architectural data[23], business process management[24],[25],[26], business modeling[27], business process re-engineering[24],[28], implementation of applications[29], context-aware systems[30], interoperability between different systems and platforms[31], knowledge management in the organization[29], and so on. Therefore, it is of paramount importance to identify a suitable ontology that has the necessary comprehensiveness, proper coverage, accuracy, compatibility, and extensibility for several applications.

This study aims to evaluate and compare enterprise ontology models from the conceptual view and then analyze their results. O'Leary[32] reviews enterprise ontology according to activity theory but doesn't consider many other aspects of enterprise ontology. Besides O'Leary's work, it could be said that there are no other proper comparisons and classifications of enterprise ontology models available. Therefore, researchers who need to use enterprise ontology models in different domains may become confused as the domain of relevance of each model is not clear.

In this paper, a framework is presented to compare enterprise ontologies at the conceptual level. Then, the criteria for the evaluation of ontology are studied, and appropriate criteria relating to the conceptual level are selected among them. Subsequently, we compare and examine enterprise ontology models using the said criteria. Finally, we analyze the results and provide a roadmap for future research on conceptual models of enterprise ontology. Section 2 reviews related works, section 3 examines conceptual models of enterprise ontology, section 4 analyzes the results of surveys, and section 5 concludes the discussion by providing suggestions for future work.

**2-Related work**

**2-1-Ontology**

According to the Gruber definition in 1993[33], an ontology is a formal, explicit specification of shared conceptualization. On the basis of this definition, "conceptualization" refers to an abstract model of phenomena in the world along with the detection of related concepts to those phenomena. "Explicit" means that the types of used concepts and their limitations are defined explicitly. "Formal" refers to the fact that the ontology should be readable to a machine and "shared "indicates that the ontology must acquire agreed and acceptable knowledge by related societies[23]. Although this definition emphasizes the formal and explicit description of concepts, these descriptions need to first agree on selected concepts and an acceptable conceptual model. If the concepts aren't chosen appropriately, the ontology usage will notbe efficient. Awell-defined conceptual model is useful in many researches and applications independently.

## 2-2-Conceptualization

A formal model is implemented in an ontology language such that the ontologist observes a gradual transition from the knowledge level to the implementation level. The formalization grade of the knowledge model increases gradually until it is able to be understood by the machine. Figure 1 shows this gradual movement.

Ontology development activities generally include: specification, conceptualization, formalization, implementation, and maintenance. Conceptualization is a crucial activity in the ontology development process[34]. Some studies emphasize this and provide methods for conceptualization. The conceptualization activity constructs meaningful knowledge models from domain knowledge. The conceptualization activity is similar to collecting puzzle pieces provided by the knowledge acquisition activity, and it is completed during the conceptualization process. The conceptualization activity must be done rigorously; otherwise, the error will propagate into the next steps.

The purpose of conceptualization is to prepare a domain model with a lower degree of formality than a formal model but still more formal than the definition of the model in natural language. Other motivations for the conceptualization process include:

1. Domain experts, human users, and ontologists may struggle to interpret or understand the ontology proposed in the ontology language.

2. Domain experts may not be able to construct ontologies within their domain of expertise.

This activity deserves special attention because it plays an important role in the ontology development process. In ontology development methodologies, after the conceptual model has been designed, the conceptual model is transformed into a formal model and implemented in an ontology language such that the ontologist observes a gradual transition from the knowledge level to the implementation level. The formalization grade of the knowledge model increases gradually until it is able to be understood by the machine. Figure 1 shows this gradual movement.

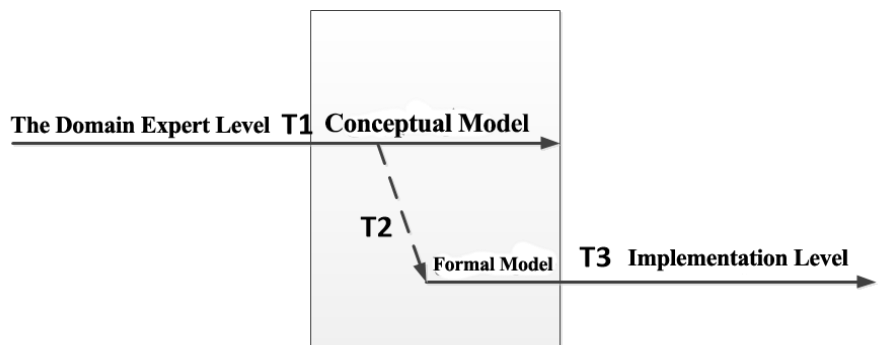

**Figure 1:** Knowledge in the Ontology Development Process [35]

T1 transformation refers to the conceptual modeling process that transforms the domain expert's subjective model into a conceptual model. T2 refers to the conceptual model progressing into a formal model. T3 refers to the formal model progressing into a model that can be understood by a machine. As the figure shows, T1 and T3 transformations are drawn by a continuous line, while the T2 transformation is indicated by a dashed line. This indicates that there may be some loss of domain knowledge when a conceptual model is developed into a formal model. This happens when the components and tools used to create conceptual models are more meaningful and expressive than those that are used to create the formal model.

In the methodology development process[36], the conceptualization activity uses a set of intermediate representations (IRs) through table and graph notation, organizing and converting an informal representation of a domain into specifications that can be understood by domain experts and ontology developers. In this study, we aim to compare enterprise ontology models from a conceptual perspective.

## 2-3- Ontology-Based Enterprise Modeling

Enterprise ontology contains a set of well-defined terms that are widely used as common descriptions of enterprises, as it accurately covers concepts related to the enterprise field. The enterprise ontology acts as an interactive medium or platform between different people, such as users, designers, and planners in various organizations[37].

An important issue in achieving integration and performing effective business planning is that all operators and stakeholders (from planning managers to low-ranking contractors involved in software production) must have a common understanding of the different dimensions of an organization; when a particular word is used from the domain, the concept it refers to should be clear. In other words, it is necessary to overcome the "semantic heterogeneity" associated with implicit perceptions of common words' meanings in the domain.

Enterprise ontology has been created for this purpose and contains a set of well-defined words that are widely used as general descriptions of an organization, covering concepts related to the domain of the organization carefully. This set facilitates a shared understanding of an organization and can serve as a fixed basis for identifying functional requirements and creating organizational models. Thus, perceptual errors are reduced in cases where the same concepts may be referred to by different terms, as it improves and facilitates interaction between stakeholders, which is an important step in increasing efficiency.

The purpose of applying an ontology in an organization is to determine the relationships between the tasks and activities of that organization with organizational knowledge and their tools. It also participates in the acquisition, representation, and manipulation of organizational knowledge, organized and structured libraries of existing knowledge, rationalized descriptions of inputs and outputs of involved components, and a vocabulary exchange format for an enterprise[38].

**3- Investigating Conceptual Models of Enterprise Ontology**

Researchers represent different ontology models for different applications. This article selects pioneering research such as TOVE[1][35], [39], context-based enterprise ontology[40], and the enterprise ontology TEO[2] [37]. These projects and other researchers are recognized as pioneers in this article. In addition, enterprise architecture frameworks require a conceptual model of an organization. Therefore, some frameworks and methodologies provide conceptual models of enterprise ontology for developing enterprise architecture, such as the Dodaf[3] Data Meta Model[41], [42],[43], Modaf[4] [44] , and the Togaf[5] Content Meta Model[45]. ArchiMate[45] is also investigated in this study. Some researchers focus only on one dimension of an organization and present an enterprise ontology model for it, as referred to in Table 1. For example, Almeida and Gizzareti[49], Pereira and Almeida[45], Santos et al. [46], Abramowicz et al. [47], and Pereira[48] have modeled organizational structure ontology and represented extensive details. Some other studies construct an enterprise ontology model for a specific type of organization. In this regard, Belo and Silva[49] have presented an enterprise ontology model specifically for higher education institutions.

---

[1] Toronto Virtual Enterprise
[2] Time Event Ontology
[3] Department of Defence Architecture Framework
[4] Ministry of Defence Architecture Framework
[5] The Open Group Architecture Framework

Table 1: Enterprise ontology which has examined the concepts of one of the dimensions of the organization

| Dimension | References |
|---|---|
| Structure | [50],[46], [51],[45],[47], [48],[52] |
| Purpose | Business Motivation Model (BMM)[53] |
| Rules and Time | Date-Time Foundation Vocabulary Request For Proposal[54] |

The best examples of such research can be found in the documentation provided by the Object Management Group (OMG). Several OMG documents serve as valuable references for analyzing the conceptual model of enterprise ontology, presenting concepts in an organized manner and detailing their relationships with one another. One notable document is the Business Motivation Model (BMM) [53], which offers a structured set of concepts that model elements of a business plan. This document is particularly useful for identifying the purpose and motivation of an organization, including concepts such as purpose, mission, perspective, and strategy. Additionally, OMG has published a comprehensive document on business process modeling known as BPMN[6]. This document not only covers conceptual control flow modeling but also provides precise definitions of components such as activities, events, gateways, and the sequences between them. It also explores the relationships between various organizational components and activities. For example, it represents capital resources using Pools and Lanes while defining consumable resources with a symbol called the Data Object. Furthermore, the Organization Structure Meta-model (OSM) [50] document from OMG offers metamodeling for organizational structures. It includes modeling elements that represent organizational entities, subgroups, features, and the relationships between organizational units and their assigned individuals. The concepts used in organizational structure are clearly defined within the OSM document. The Semantics of Business Vocabulary and Rules (SBVR) [55] takes this a step further by expressing business vocabularies with formal logic, providing a specific language for business descriptions. It defines a set of terms, each with a specific technical meaning relevant to the field of business. The rules defined by formal logic closely resemble natural language, making SBVR a business ontology that machines can understand[55]. Lastly, the Date Time Foundation

---

[6] [Https: //www. omg. org/spec/BPMN/2. 0/PDF]

Vocabulary Request for Proposal[54] is another OMG document that articulates concepts related to time and dates in business using SBVR.

**3-1- Providing a Framework for Comparison**

There has been no existing framework to compare the conceptual models of enterprise ontology until now, making our study the first to address this issue. In this regard, we present a framework for evaluating conceptual models of enterprises. Two perspectives have been considered in formulating this framework, which includes various comparison parameters. The first perspective aims to identify the closest semantic frameworks to enterprise ontology and draws inspiration from their parameters for comparison. The second perspective identifies and applies general parameters for evaluating ontologies, supported by detailed research in the field. In the first perspective, the most significant research was conducted by Osterwalder[56], who presented a framework for comparing business model ontologies based on earlier works[57],[58]. Our study generalizes Osterwalder's framework to facilitate the comparison of enterprise ontologies. This generalization seeks to provide an acceptable framework for comparing ontologies that model organizations. In this framework, we introduce important parameters for the comparison of enterprise ontology, which are described as follows:

1. **Purpose**: This parameter reflects the primary motivation behind enterprise ontology and the objectives of its development. Several studies, such as[41], present ontology models specifically for applications in enterprise architecture. Others[59], like , concentrate on ontology models tailored for business contexts. Additionally, some studies, such as[40], offer a more general enterprise ontology aimed at enterprise modeling.
2. **Domain**: This parameter evaluates the domain of relevance for the conceptual model. The enterprise ontology model is capable of representing various types of enterprises, including business, military, and educational organizations.
3. **Implementation language**: This parameter indicates the programming languages used to implement enterprise ontology and convert it into a machine-readable format. It encompasses the use of generic ontological technologies for representing ontologies, such as Ontolingua, RDF/S, and OWL, as well as ontology design tools like Protégé.

4. **Representation**: Lightweight ontologies encompass concepts, classifications of those concepts, the relationships between them, and characteristics that describe each concept. In contrast, heavyweight ontologies build upon lightweight ontologies by adding axioms and constraints that clarify the relationships within the collected vocabularies. Heavyweight ontologies are typically implemented using approaches based on artificial intelligence, employing first-order logic or description logic. On the other hand, lightweight ontologies are represented through software engineering methods, such as UML or database diagrams like ERD.

5. **Ontology content and components:** This parameter refers to the key dimensions addressed by each enterprise ontology. Each conceptual model of an organization encompasses various dimensions and the concepts associated with them. It is crucial to identify which dimensions are included at the macro level and which concepts are most critical among all the concepts. Additionally, this parameter takes into account the types of relationships and the nature of the rules present within the ontology.

6. **Ontology maturity and evaluation:** The degree of maturity of an ontology is determined by its evaluation. Various qualitative criteria and resources exist for assessing ontological models that have been implemented, as detailed in references such as [60], [61], [62] and . Several criteria possess significant capability for studying the evaluation of conceptual models. In this article, we select specific criteria to understand and compare the existing ontological models of organizations. If the evaluation criteria function effectively at the conceptual level, we can anticipate that the enterprise ontology will perform well at the formal level. The selected criteria for evaluation are as follows:
    a. **Reusability**: Reusability refers to the transferability of an ontology, specifically which parts can be utilized to construct another ontology for a different purpose. Given that ontology design and implementation can be challenging and time-consuming, reusability is a crucial factor.

b. **Accuracy**: Accuracy requires that the underlying knowledge informing the ontology aligns with the expertise of domain specialists. Ontology models must accurately describe the real world, despite the inherent lack of precise semantics that allows for various interpretations. Therefore, an organizational ontology must demonstrate strong alignment with key entities and correspond well with our understanding of the organization.

  c. **Expandability**: Expandability refers to the ability to extend the ontology into other domains without changing its definitions.

  d. **Adaptability**: Adaptability reflects how well the ontology can anticipate future developments. It assesses whether the ontology provides a solid foundation that is easily expandable and sufficiently flexible to respond predictably to minor internal changes. Unfortunately, many ontologies do not offer an adequate basis for future expansion.

  e. **Completeness**: Completeness examines whether the model has sufficient domain coverage to enable the ontology to answer all relevant questions within that domain.

The parameters are summarized in Table 2. This table, along with its parameters, provides a framework for comparing enterprise ontologies, which we will discuss in this section.

**Table 2:** Ontology Evaluation Parameters That Are More Aligned with Conceptual Model Evaluation

| Parameter | | References | Description |
|---|---|---|---|
| Purpose | | [56] | The main motivation of organizational ontology and the purpose of creating the ontology. |
| Domain | | [56] | The domain of organization which the ontology is modeling, for example business, military or educational organization. |
| Implementation language | | [56] | The implementation language and applied language used to create the enterprise ontology. |
| Representation | | [56] | How the the enterprise ontology model is represented. |
| Ontology content and component | | [56] | The dimensions of the domain considered by the conceptual model and the concepts underpinning them. |
| Ontology maturity and evaluation | Reusability | [60] | The degree to which the entire ontology or part thereof can be repurposed and reconstructed another ontology. |
| | Accuracy | [7] | The degree of consistentency of the ontology with the knowledge of a domain expert. |
| | Expandability | [34] | The ability to extend the ontology to other domains in without changing definitions. |
| | Adaptability | [60] | Whether the model reacts predictably towards the small internal changes or not. |
| | Completeness | [7] | The ability to how exhaustively the ontology as answer all questions that ontology should be able to answer. |

### 3-2- Comparison of Enterprise Ontologies

In this section, we compare conceptual models according to our framework presented in the previous section.

**Purpose**: The main motivation of organizational ontology and the purpose of creating the ontology.The number of studies such as TOGAF Content model[63] [64] , ArchiMate [45],[65] DODAF Data Meta Model[41], UAF[7][66],[67] provided ontology models for enterprise architecture applications. Some studies such as context-based[40] and The Enterprise Ontology(TEO) [37] and TOVE[68] presented enterprise ontology in general for enterprise modeling.

**Domain**: The domain refers to the specific type of organization that the ontology is designed to model. This could encompass various sectors such as business, military, or educational organizations. Each domain has its unique characteristics, structures, and processes, which the ontology aims to represent accurately. By tailoring the ontology to a particular domain, it becomes more relevant and useful for stakeholders within that field, facilitating better understanding, communication, and decision-making.

TOGAF Content Model[63] [64], ArchiMate[65], and the Unified Architecture Framework (UAF) [66],[67] are generic frameworks in the realm of enterprise ontology. In contrast, other models, such as[59] presented in , focus specifically on ontology models for the business domain. Additionally, the DODAF Data Meta Model[41] provides an ontology model tailored for the military domain .

**Implementation language**: Most ontology models such as Dodaf Data Meta Model [41], Togaf Content Model[63] [64], ArchiMate[65], UAF[66],[67] are represented at the conceptual level by UML diagrams. TOVE[68] implemented by Prolog language and The Enterprise Ontology(TEO) [37] implemented by Ontolingua language (based on KIF).

Most ontology models, including the DODAF Data Meta Model, TOGAF Content Model, ArchiMate, and UAF, are typically represented at the conceptual level using UML diagrams. In contrast, TOVE is implemented using the Prolog programming language, while The Enterprise Ontology (TEO) is developed using Ontolingua, which is based on KIF.

**Representation**: Most ontology models are implemented in a lightweight form, while only TOVE and The Enterprise Ontology (TEO) [37] are classified as heavyweight ontology models.

**Content and component**: The core conceptual model identifies the main dimensions of an organization, followed by the detailed concepts that support each dimension. For example, The Enterprise Ontology (TEO) [37] represents five key dimensions: activity, organization, strategy, marketing, and time.

---

[7] Unified Architecture Framework

**Reusability:** Most enterprise ontologies immediately transition into the implementation phase without first establishing a solid conceptual model. This oversight limits users' ability to connect the abstract concepts of the model with the real-world elements they are intended to represent. A robust conceptual model is crucial for supporting reusability. Among organizational ontology models, both TOVE and The Enterprise Ontology (TEO) rush into implementation, leaving users without a comprehensive understanding of the models, which hampers effective usage. In contrast, the DODAF Data Meta Model offers a well-defined conceptual model that articulates relationships at a conceptual level, although it is specifically tailored for the United States Department of Defense. The Context-Based Enterprise Ontology starts with a conceptual level presentation, but many of its relationships remain unclear, limiting its reusability. Meanwhile, the Unified Architecture Framework (UAF) describes each concept precisely, supporting extensibility; however, it suffers from ambiguities at the macro level, making reusability challenging.

**Accuracy:** TOVE and TEO exhibit ambiguous concepts, with their definitions and relationships not being clearly defined. This leads to varying interpretations of each concept. In contrast, the DODAF Data Meta Model provides a well-defined enterprise ontology aimed at representing the conceptual model of defense organizations. Additionally, the UAF, which originated from DODAF and MODAF, is designed to support non-defense organizations. The TOGAF Content Meta Model clearly defines concepts and their relationships, offering a solid foundation for enterprise ontology criteria; however, it does not implement the ontology model at a formal level.

**Expandability**: Context-based enterprise ontology, TOGAF Content Meta Model, ArchiMate, and UAF are generally defined in a way that allows for good expandability into specific domains. In contrast, the relationship between "activity" and "capability" in the DODAF Data Meta Model is tailored specifically for military organizations, which limits its applicability to other sectors despite its well-defined concepts.

**Adaptability**: TOVE and TEO transition abruptly into the formal phase without adequately defining their concepts. On the other hand, the DODAF Data Meta Model excels in defining the conceptual phase but is primarily suited for military organizations. The TOGAF Meta-Model considers strong concepts at the micro level, yet it lacks a comprehensive ontological structure.

**Completeness**: Completeness refers to how thoroughly an ontology can address all relevant questions about the organization it represents. This means that the ontology should cover all dimensions of an organization. Given the social nature of organizations, researchers must consider multiple dimensions simultaneously. For instance, defining "service" requires acknowledging the roles of both service customers and providers. Similarly, a complete understanding of a "business process" necessitates describing the roles of its participants. To achieve this, the structure of organizational units and the roles within them must be framed within a broader organizational context. However, focusing on too many dimensions can lead to selecting concepts that may not be essential for a complete description of the domain. An ontology can be considered comprehensive if it effectively answers questions such as: Who (performer) does what (task) for what reasons (goal), where (location), and when (time) [40]?

**Table 2:** Comparison of Enterprise Ontology According to the Proposed Framework

|  | TOVE | The Enterprise Ontology(TEO) | Context-based | DODAF Data Meta Model | TOGAF Content Model | ArchiMate | UAF |
|---|---|---|---|---|---|---|---|
|  | [68] | [37] | [40] | [41] | [64], [63] | [45], [65] | [66], [67] |
| Purpose | Enterprise modeling | Enterprise modeling | Enterprise modeling | Enterprise Architecture | Enterprise Architecture | Enterprise Architecture | Enterprise Architecture |
| Domain | Public and commercial | Commercial Enterprise | Public | Military | Public | Public | Public |
| Implementation language | Prolog | Ontolingua (Base on KIF) | At the conceptual level and UML | At the conceptual level and UML | At the conceptual level and UML | has provided its own modeling language | At the conceptual level and UML |
| Representation | Heavyweight | Heavyweight | lightweight | lightweight | lightweight | lightweight | lightweight |
| Ontology content and components | Organization, Resource, | Organization's Activity, | Purpose area, Actor area, | Activity, Capability, Resource | Governance, Service, Process, | 3 layers of business, applicati | Taxonomy Structure Connecti |

|  |  | TOVE | The Enterprise Ontology(TEO) | Context-based | DODAF Data Meta Model | TOGAF Content Model | ArchiMate | UAF |
|---|---|---|---|---|---|---|---|---|
|  |  | [68] | [37] | [40] | [41] | [64], [63] | [45], [65] | [66], [67] |
|  |  | Activity, time, cost | Strategy, Marketing, Time | Action area, Object area, Facility area, LOcation area, Time area | (Information, Performer and Material), Location, Guide | Data, Infrastructure, Motivation | on and technology that stand under each concepts. | vity Processes States Interaction Scenarios Information Constraints Roadmap Traceability |
| Ontology maturity and evaluation | Reusable | Mid | Mid | High | Mid | Mid | Mid | Mid |
|  | Accuracy | Low | Low | Mid | Mid | High | High | Mid |
|  | Expandability | Low | Low | High | High | High | High | High |
|  | Adaptability | Low | Low | Low | Mid | High | High | High |
|  | Completeness | Low | Low | Mid | High | High | High | High |

## 3-3- Investigating the Ontology of Enterprise Completeness Using the Zachman Framework

Enterprise ontology studies examine various dimensions of organizations. For instance, Leppanen introduced a context-based ontology[40] that encompasses seven dimensions: goal, actor, action, object, facility, location, and time. The TOVE project[68], [39] considered four dimensions: ontology of organization[68], ontology of resource[69], ontology of activity[70], and ontology of cost[71]. ArchiMate[65] presented a meta-model structured into three layers: business, application, and technology. Additionally, ArchiMate categorizes its elements into three groups: active elements that perform actions, behavioral elements that represent the behavior of active elements, and passive elements that are acted upon by behavioral elements. The TOGAF Content Model[64], [63] outlines concepts such as motivation, infrastructure, data, process, service, and governance at the first level.

The Zachman Framework[72], [73] aims to examine all dimensions of an organization, making it a suitable foundation for studying enterprise ontology models. The columns of the Zachman Framework represent various aspects (dimensions) of an organization, derived from the 5W1H questions: who (responsibilities), when (time), why (motivation), where (location), how (task), and what (data). A central question arises: does the ontology cover all dimensions of the enterprise? If it does, then the enterprise ontology will demonstrate good comprehensiveness. The six communication questions of 5W1H help clarify the dimensions of organizations, as noted by Caetano et al. [74] and Zachman. Rajabi et al. [17] present an enterprise ontology model based on the factors in the columns of the Zachman Framework. By utilizing the 5W1H questions, Zachman clarifies each dimension of the organization, providing a solid basis for understanding existing ontologies and their covered dimensions. In this paper, we will compare the completeness of enterprise ontology models using the columns of the Zachman Framework.

In addition to reviewing enterprise ontology concepts (see Table 3), we also assess their compatibility with the Zachman Framework. The selected ontologies for comparison are leading models that have sufficient documentation available. For instance, the TOGAF Content Meta-Model introduces important concepts such as data entity, value stream, constraint, role, organization unit, location, business service, process, function, and business capability. Each of these concepts is comprehensively defined along with their relationships to one another. Within the Zachman Framework, the concepts can be categorized as follows: data entity and value stream fall under the "What" column; process, function, and business service are placed in the "How" column; location is categorized under "Where"; and organization unit is found in the "Who" column. Notably, there are no concepts represented under the "When" column. As illustrated in Table 3, the Context-Based Enterprise Ontology defines specific areas for each column of the Zachman Framework and outlines various concepts for each area. This model demonstrates better alignment with the columns of the Zachman Framework, enhancing its adaptability and relevance.

Some ontology models include "business product" as a key concept within enterprise ontology. For instance, ArchiMate[65],[65] considers "product" as an essential element of the organization, defining it as anything offered to the outside world. This definition also encompasses products that may be provided internally to different parts of the organization. Thus, the concept of "product" is crucial, yet it is overlooked in some other ontologies.

The concept of "location" is included in the ontology of organizations, but some models, such as DODAF, TOGAF, and ArchiMate, limit their definition to a single concept of "location." In contrast, the Context-Based Enterprise Ontology and the Unified Architecture Framework (UAF) consider multiple concepts related to location. On the other hand, both TOVE and enterprise ontology do not address any concepts for the "where" column. The concept of "business service" in TOGAF[64] supports business capabilities through an explicitly defined interface and is governed by an organization. Similarly, the UAF defines "service specification" as a set of functionalities provided by one element for use by others. This indicates that the concept of "service" is significant within enterprise architecture; however, it is absent from models that focus solely on organizational modeling. Table 4 summarizes these findings regarding adaptability within the Zachman Framework, while Table 5 presents the final results for the adaptability of existing models according to the Zachman Framework.

**Table 4:** A Review of the Adaptability of Existing Concepts in Enterprise Ontology Within the Zachman Framework

| Enterprise ontology | Why | What | How | Who | Where | When |
|---|---|---|---|---|---|---|
| TOVE[68] | | Goal, Sub goal | Activity, Constraint, Authority Communication link | Resource, Organization, Division, Subdivision, Team, Agent, Role, Skill | | |
| The Enterprise ontology [37] | Purpose, Mission | CSF, Objective, Vision, Goal | Activity, Activity Spec, Sub-Activity, Execute, Plan, Sub-Plan, Process Spec, Org. Structure, Strategy, Risk, Capability | Entity, Role, Relation Attribute Resource, Person, Corporation, Unit, Actor, Machine, Actor Role, Skill, Activity Owner, Doer, Authority | | Time point, Time Interval, T-Begin, T-End, Time Line, Calendar, Date, Duration |
| Context-based Enterprise Ontology [40] | Reason, Purpose | Goal | Function, Activity, Task, Action structure | Facility, Resource, Tool, Human actor, Person, Group, Position, role, Unit, Organization | Location Area, Physical location Point, Spatial thing Logical, location Region, Geographical dimension, Geographical system | Time, Time point, Time interval, Time unit, Time system, Clock time, Calendar time |
| DODAF Data Meta Model [41] | | Vision, Desired Effect | Project, Capability, Activity, Guidance, Condition | Personnel Type, Skill, Performer, Data, Information, Materiel | Location | |
| TOGAF Content Meta Model [64], [63] | | Business Service | Function, Process, Value Stream, Course of Action, Business Capability, Constraint | Organization Unit, Function, Role, Data Entity | Location | |
| ArchiMate[45],[65] | | Business service, Business product | Business process, Business function, Business interaction, Business event, Contract | Business Role, Business actor, Business collaboration, Business object, | Location, Business interface | |
| UAF [66],[67] | | Service, Service Specification, EnterpriseVision, EnterpriseGoal | Capability, Project Kind, Project Activity, Project Milestone, Capable Element Project, Actual Milestone Kind, Operational Activity | Actual Organization, Organizational Resource, Person, Post, Responsibility, Natural Resource, Physical Resource, Resource Architecture, Resource Artifact, Resource Performer, Software System, Standard, Protocol, Protocol Stack | Location, Location Holder, Location Kind, Actual Location | |

Table 5: Results of the Adaptability of Existing Concepts in Enterprise Ontology Within the Zachman Framework

| Conceptual Model | TOVE | The Enterprise ontology(TEO) | Context-based | DODAF Data Meta Model | TOGAF Content model | ArchiMate | UAF |
|---|---|---|---|---|---|---|---|
| Adaptability with the Zachman framework | × | × | ✓ | × | × | × | × |

## 4- Results analysis

Numerous studies have explored enterprise ontology; however, many exhibit significant weaknesses. A critical issue is the lack of consensus on which dimensions of an organization should be included in its ontological model. For instance, while one study[68] incorporates the time dimension, another study[41] overlooks it entirely. Establishing a common agreement on the conceptual model is essential before progressing to formal and logical construction phases. We require a foundational conceptual model that accurately represents key concepts and relationships, yet currently, there are no standard or reference models available for researchers to consult.

The existing enterprise ontology models often lack conceptual depth, leading to premature transitions into the implementation phase. This results in underdeveloped conceptual models, making their formal counterparts neither reusable nor expandable. Furthermore, comprehending formal models becomes challenging without a robust conceptual framework.

Additionally, the concepts present in current ontologies do not adequately encompass all components of an organization. If ontological models were better aligned with the columns outlined in Zachman's framework, they could more effectively cover various organizational dimensions. There is an urgent need for a powerful conceptual model that articulates fundamental concepts and relationships in an interpretable manner for diverse applications, including enterprise architecture, business architecture, business process management, context-aware systems, intercommunication, and automated production and analysis of models.

Moreover, there is no standardized approach to expand and customize a generic enterprise ontology model for specific domains or organizational needs.

Conceptual models of enterprises can be enriched progressively: initially, a general enterprise ontology model is developed; then it is refined for specific industries; ultimately, an enterprise-specific ontology emerges based on previous models. This progression is illustrated in Figure 2. In general, all organizations share a set of common principles and concepts that form their foundation. In the second stage, these common concepts are detailed to suit various types of organizations—such as commercial entities, military organizations, and universities. Finally, in the third step, appropriate common concepts are tailored specifically to describe individual organizations

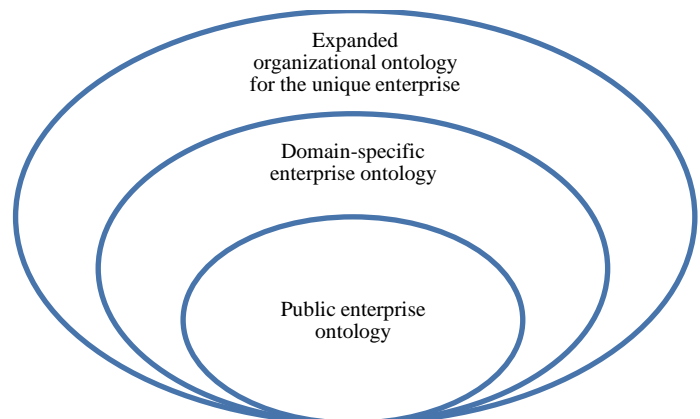

**Figure 2:** The Relationship Between Enterprise Ontology and Its Subsets (Development and Expansion of Ontology)

## 5- Conclusion

Enterprise ontology offers a comprehensive and systematic framework that enhances the understanding of organizations for both managers and stakeholders. By addressing ambiguities and contradictions, this framework empowers informed decision-making. The clarity it provides is invaluable not only for human users but also for machine processing. This structured understanding is applicable across various domains, including efficient modeling of enterprise architecture, business processes, and context-aware systems. In this study, we examined and compared conceptual models of enterprise ontology, highlighting their strengths and weaknesses. Future research should focus on developing a reference enterprise ontology model that encompasses all dimensions of an organization, ensuring it covers every component while identifying key elements and relationships. Additionally, there is a critical need to establish reliable methods for adapting the reference enterprise ontology model to create tailored models for specific domains, organizations, or applications.

**Conflicts of Interests**

The authors did not receive support from any organization for the submitted work.

**Author Contributions Statement**

Seyed Mohsen Rahnamafard conceptualized the research idea, laying the groundwork for the study. Zeinab Rajabi further developed this concept, refining the initial ideas and enhancing the framework for analysis. Both authors, collaborated closely in conducting the analysis and comparisons presented in this study, ensuring a comprehensive evaluation of the existing literature on enterprise ontology.

**References**

1. Daft, R.L., *Organization theory & design*. 2020: Cengage learning.

2. Stacey, R.D., D. Griffin, and P. Shaw, *Complexity and management: Fad or radical challenge to systems thinking?* 2000: Psychology Press.

3. Mingers, J., *Realising systems thinking: knowledge and action in management science*. 2006: Springer Science & Business Media.

4. Robbins, S.P., *Organization Theory: Structures, Designs, And Applications, 3/e*. 1990: Pearson Education India.

5. Senge, P.M., *The art and practice of the learning organization*. 1990, New York: Doubleday.

6. Porras, J.I. and P.J. Robertson, *Organization development theory: A typology and evaluation*. 1986: Graduate School of Business, Stanford University.

7. Burke, W.W. and G.H. Litwin, *A causal model of organizational performance and change.* Journal of management, 1992. **18**(3): p. 523-545.

8. Nadler, D.A. and M.L. Tushman, *Types of organizational change: From incremental improvement to discontinuous transformation.* Discontinuous change: Leading organizational transformation, 1995: p. 15-34.

9. Weisbord, M.R., *Organizational diagnosis: Six places to look for trouble with or without a theory.* Group & Organization Studies, 1976. **1**(4): p. 430-447.

10. Waterman Jr, R.H., T.J. Peters, and J.R. Phillips, *Structure is not organization.* Business horizons, 1980. **23**(3): p. 14-26.

11. Levitt, H.J., *Applied organizational change in industry: Structural, technological, and humanistic approach.* Handbook of Organizations, 1965. **1144**: p. 1170.

12. Ghani, A.S.D.U., *Application Of Weisbord's Organizational Diagnosis Model A Case of Pakistan Banking Industry.* 2013.

13. Janićijević, N., *Business processes in organizational diagnosis.* Management: journal of contemporary management issues, 2010. **15**(2): p. 85-106.

14. Rosing, M.v., *Overview of the Business Ontology Research & Analysis*. 2015.

15. Dietz, J.L., *What is Enterprise Ontology?* 2006: Springer.

16. zur Muehlen, M., *Enterprise Architecture based on Design Primitives and Patterns-Guide for the Design and Development of Event-Trace Descriptions (DoDAF OV-6c) using BPMN.* Business Transformation Agency of United States Department of Defense, Version, 2009. **1**.

17. Rajabi, Z., B. Minaei, and M.A. Seyyedi, *Enterprise architecture development based on enterprise ontology.* Journal of theoretical and applied electronic commerce research, 2013. **8**(2): p. 85-95.

18. Kindrick, J., *Enterprise Architecture based on Design Primitives and Patterns business transformation agency.* Availaible on: http://www. bta. mil, 2009.

19. Rajabi, Z. and M.N. Abade, *Data-Centric Enterprise Architecture.* International Journal of Information Engineering and Electronic Business, 2012. **4**(4): p. 53.

20. Model, M., *The DoDAF Architecture Framework Version 2.02*.

21. Feilmayr, C. and W. Wöß, *An analysis of ontologies and their success factors for application to business.* Data & Knowledge Engineering, 2016. **101**: p. 1-23.

22. Hinkelmann, K., et al., *A new paradigm for the continuous alignment of business and IT: Combining enterprise architecture modelling and enterprise ontology.* Computers in Industry, 2016. **79**: p. 77-86.

23. Antunes, G., et al., *Using Ontologies for Enterprise Architecture Integration and Analysis.* CSIMQ, 2014. **1**: p. 1-23.

24. Rao, L., G. Mansingh, and K.-M. Osei-Bryson, *Building ontology based knowledge maps to assist business process re-engineering.* Decision Support Systems, 2012. **52**(3): p. 577-589.

25. Santos Jr, P.S., J.P.A. Almeida, and G. Guizzardi. *An ontology-based semantic foundation for ARIS EPCs.* in *Proceedings of the 2010 ACM Symposium on Applied Computing.* 2010. ACM.

26. Jung, J.J., *Semantic business process integration based on ontology alignment.* Expert Systems with Applications, 2009. **36**(8): p. 11013-11020.

27. Gassen, J.B., et al., *An experiment on an ontology-based support approach for process modeling.* Information and Software Technology, 2017. **83**: p. 94-115.

28. AbdEllatif, M., M.S. Farhan, and N.S. Shehata, *Overcoming business process reengineering obstacles using ontology-based knowledge map methodology.* Future Computing and Informatics Journal, 2018. **3**(1): p. 7-28.

29. Villela, K., et al., *The use of an enterprise ontology to support knowledge management in software development environments.* Journal of the Brazilian Computer Society, 2005. **11**(2): p. 45-59.

30. Aguilar, J., M. Jerez, and T. Rodríguez, *CAMeOnto: Context awareness meta ontology modeling.* Applied computing and informatics, 2018. **14**(2): p. 202-213.

31. Chen, D., G. Doumeingts, and F. Vernadat, *Architectures for enterprise integration and interoperability: Past, present and future.* Computers in industry, 2008. **59**(7): p. 647-659.

32. O'Leary, D.E., *Enterprise ontologies: Review and an activity theory approach.* International Journal of Accounting Information Systems, 2010. **11**(4): p. 336-352.

33. Gruber, T.R., *Toward principles for the design of ontologies used for knowledge sharing?* International journal of human-computer studies, 1995. **43**(5-6): p. 907-928.

34. Gómez-Pérez, A., *Towards a framework to verify knowledge sharing technology.* Expert Systems with applications, 1996. **11**(4): p. 519-529.

35. Gomez-Perez, A., M. Fernández-López, and O. Corcho, *Ontological Engineering: with examples from the areas of Knowledge Management, e-Commerce and the Semantic Web.* 2006: Springer Science & Business Media.

36. Fernández-López, M., A. Gómez-Pérez, and N. Juristo, *Methontology: from ontological art towards ontological engineering.* 1997.

37. Uschold, M., et al., *The enterprise ontology.* The knowledge engineering review, 1998. **13**(1): p. 31-89.

38. Ciocoiu, M., D.S. Nau, and M. Gruninger, *Ontologies for integrating engineering applications.* Journal of Computing and Information Science in Engineering, 2001. **1**(1): p. 12-22.

39. Fox, M.S., M. Barbuceanu, and M. Gruninger. *An organisation ontology for enterprise modelling: preliminary concepts for linking structure and behaviour*. in *Enabling Technologies: Infrastructure for Collaborative Enterprises, 1995., Proceedings of the Fourth Workshop on*. 1995. IEEE.

40. Leppänen, M. *A context-based enterprise ontology*. in *International Conference on Business Information Systems*. 2007. Springer.

41. officer, C.i., *DoDAF Architecture Framework Version 2.02 - DoD CIO*. 2009.

42. Thakor, P. and S. Sasi, *Ontology-based Sentiment Analysis Process for Social Media Content.* Procedia Computer Science, 2015. **53**: p. 199-207.

43. *DoDAF PLUGIN user guide*. 2010. **version 16.9**.

44. Aue, A. and M. Gamon. *Customizing sentiment classifiers to new domains: A case study*. in *Proceedings of recent advances in natural language processing (RANLP)*. 2005.

45. Pereira, D.C. and J.P.A. Almeida. *Representing Organizational Structures in an Enterprise Architecture Language*. in *FOMI@ FOIS*. 2014.

46. Santos Jr, P.S., J.P.A. Almeida, and G. Guizzardi, *An ontology-based analysis and semantics for organizational structure modeling in the ARIS method.* Information Systems, 2013. **38**(5): p. 690-708.

47. Abramowicz, W., et al. *Organization structure description for the needs of semantic business process management*. in *3rd international Workshop on Semantic Business Process Management colocated with 5th European Semantic Web Conference*. 2008.

48. Pereira, D.C., *Representing organizational structures in enterprise architecture: An ontology-based approach*. 2015, Universidade Federal do Espírito Santo.

49. Silva, C. and O. Belo. *A Core Ontology for Brazilian Higher Education Institutions*. in *CSEDU (2)*. 2018.

50. OMG, *Organization Structure Metamodel(OSM)*. 2009, Object Mangement Group.

51. Almeida, J.P.A. and G. Guizzardi. *A semantic foundation for role-related concepts in enterprise modelling*. in *Enterprise Distributed Object Computing Conference, 2008. EDOC'08. 12th International IEEE*. 2008. IEEE.

52. Carvalho, V.A. and J.P.A. Almeida. *A semantic foundation for organizational structures: a multi-level approach*. in *2015 IEEE 19th International Enterprise Distributed Object Computing Conference*. 2015. IEEE.

53. OMG, *Business Motivation Model (BMM)*. 2015, Object Mangement Group.

54. OMG, *Date-Time Foundation Vocabulary Request For Proposal*. 2008, Object Management Group.

55. Kang, D., et al., *An ontology-based enterprise architecture.* Expert Systems with Applications, 2010. **37**(2): p. 1456-1464.

56. Gordijn, J., A. Osterwalder, and Y. Pigneur, *Comparing two business model ontologies for designing e-business models and value constellations.* BLED 2005 Proceedings, 2005: p. 15.

57. Jasper, R. and M. Uschold. *A framework for understanding and classifying ontology applications*. in *Proceedings 12th Int. Workshop on Knowledge Acquisition, Modelling, and Management KAW*. 1999.

58. Pateli, A., *A framework for understanding and analysing ebusiness models.* BLED 2003 Proceedings, 2003: p. 4.

59. Poels, G., et al. *Conceptualizing Business Process Maps*. arXiv e-prints, 2018.

60. McDaniel, M., V.C. Storey, and V. Sugumaran, *Assessing the quality of domain ontologies: Metrics and an automated ranking system.* Data & Knowledge Engineering, 2018. **115**: p. 32-47.

61. Brank, J., M. Grobelnik, and D. Mladenić, *A survey of ontology evaluation techniques.* 2005.

62. Hlomani, H. and D. Stacey, *Approaches, methods, metrics, measures, and subjectivity in ontology evaluation: A survey.* Semantic Web Journal, 2014. **1**(5): p. 1-11.

63. Weisman, R., *An Overview of TOGAF®Version 9.1*. 2011.

64. Awadallah, R., *Methods for constructing an opinion network for politically controversial topics*. 2013.

65. Wierda, G., *Mastering ArchiMate Edition III: A serious introduction to the ArchiMate enterprise architecture modeling language*. 2017: R&A.

66. OMG, *Unified Architecture Framework Profile (UAFP)*. 2017, Object Management Group.

67. OMG, *Unified Architecture Framework (UAF) The Domain Metamodel Version 1.0 - Appendix A*. 2017, Object Management Group.

68. Fox, M.S. and M. Gruninger, *Enterprise modeling.* AI magazine, 1998. **19**(3): p. 109.

69. Fadel, F.G., M.S. Fox, and M. Gruninger. *A generic enterprise resource ontology*. in *Enabling Technologies: Infrastructure for Collaborative Enterprises, 1994. Proceedings., Third Workshop on*. 1994. IEEE.

70. Gruninger, M. and M.S. Fox, *An activity ontology for enterprise modelling.* Submitted to AAAI-94, Dept. of Industrial Engineering, University of Toronto, 1994. **321**.

71. Tham, K.D., M.S. Fox, and M. Gruninger. *A cost ontology for enterprise modelling*. in *Proceedings of 3rd IEEE Workshop on Enabling Technologies: Infrastructure for Collaborative Enterprises*. 1994. IEEE.

72. Zachman, J.A., *A framework for information systems architecture.* IBM systems journal, 1987. **26**(3): p. 276-292.

73. Sowa, J.F. and J.A. Zachman, *Extending and formalizing the framework for information systems architecture.* IBM systems journal, 1992. **31**(3): p. 590-616.

74. Caetano, A., C. Pereira, and P. Sousa, *Generation of business process model views.* Procedia Technology, 2012. **5**: p. 378-387.